\documentclass[aps,pra,twocolumn]{revtex4-1} 
\usepackage{graphicx,amssymb,amsfonts,amsmath}

\usepackage{hyperref,color}
\usepackage{amsmath,mathtools}
\usepackage{bbold}
\usepackage[utf8]{inputenc} 
\usepackage{graphicx}
\usepackage{braket}
\usepackage{ragged2e}
\usepackage{wrapfig}

\newcommand{\be}{\begin{eqnarray}}
\newcommand{\ee}{\end{eqnarray}}

\newcommand{\bp}{\begin{pmatrix}}
\newcommand{\ep}{\end{pmatrix}}

\newcommand{\nocontentsline}[3]{}
\newcommand{\tocless}[2]{\bgroup\let\addcontentsline=\nocontentsline#1{#2}\egroup}

\begin{document}

\title{Beyond Strong Coupling in a Massively Multimode Cavity}
\author{Neereja M. Sundaresan }
\author{Yanbing Liu}
\author{Darius Sadri}
\author{L\'{a}szl\'{o} J. Sz\H{o}cs}
\author{Devin L. Underwood}
\author{Moein Malekakhlagh }
\author{Hakan E. T\"{u}reci}
\author{Andrew A. Houck \thanks{aah@princeton.edu}}
\affiliation{Department of Electrical Engineering, Princeton University, Princeton, New Jersey 08544, USA}

%
%
%
%
%
%
%
%


\justify

\begin{abstract}

The study of light-matter interaction has seen a resurgence in recent years, stimulated by highly controllable, precise, and modular experiments in cavity quantum
electrodynamics (QED) \cite{haroche2006exploring}.  The achievement of strong coupling \cite{PhysRevLett.68.1132,Wallraff2004,Reithmaier2004}, where the coupling between a single atom and fundamental cavity mode exceeds the decay rates, was a major milestone that opened the doors to a multitude of new
investigations \cite{Mabuchi15112002, Schoelkopf2008}.
Here we investigate multimode strong coupling (MMSC) \cite{Egger2013,Krimer2014}, where the coupling is comparable to the free spectral range (FSR) of the cavity, i.e. the rate at which a qubit can absorb a photon from the cavity is comparable to the round trip transit rate of a photon in the cavity. We realize, via the circuit QED architecture \cite{Blais2004, Devoret2013}, the first experiment accessing the MMSC regime, and report remarkably widespread and structured resonance fluorescence, whose origin extends beyond cavity enhancement of sidebands \cite{Kim2014}. Our results capture complex multimode, multiphoton processes, and the emergence of ultranarrow linewidths. Beyond the novel phenomena presented here, MMSC opens a major new direction in the exploration of light-matter interactions.

\end{abstract}

\maketitle

Interest in going beyond strong coupling has focused on the ultrastrong coupling limit, where the breakdown of the rotating-wave approximation results in excitation non-conserving terms \cite{Niemczyk2010}. In contrast, the direction which we pursue is the simultaneous strong coupling of the qubit to numerous modes, leading to qubit mediated mode-mode interactions and many-body physics not present in the single mode problem. Unlike the spin-boson problem, where the continuum bosonic modes are treated as a bath for the qubit, in MMSC the dynamics of individual modes and finite time correlations between modes are
essential. Furthermore, the study of MMSC can not rely on the integrability present in the single mode and continuum problems \cite{Braak2011}, thus requiring new theoretical ideas for its study.

The closed system is described by the Hamiltonian:
\begin{equation}
H = \frac{\hbar}{2}\omega_a \sigma_z  + \sum_{m} \hbar\omega_m a^{\dagger}_m a_m +\hbar g_m( \sigma^{+} + \sigma^{-})(a^{\dagger}_m+a_m)
\end{equation}
where $\omega_a$ is the qubit frequency, $\sigma_z$ is the Pauli operator, and $m$ represents cavity mode number. Here, $a_m^{\dagger}$ ($\sigma^+$) and $a_m$ ($\sigma^-$) are mode (qubit) raising and lowering operators. The coupling strength of the qubit to the $m$th harmonic is $g_{m} = g_0\sqrt{m+1}$, where $g_0$ is the coupling rate to the fundamental cavity mode \cite{Houck2008}. 

To achieve MMSC, the coupling between qubit and cavity must be comparable to the free spectral range, which is made possible through use of a long cavity.
Previously, long cavities have been used for novel comb generation, with a Kerr nonlinearity provided by the bulk medium \cite{Erickson2014}. 
In cavity QED, unique multimode platforms have been studied, where degenerate modes are coupled via ensembles of atoms \cite{Gopala2010,Wickenbrock2013}.
In contrast, here the nonlinearity is provided by a single qubit.
In our setup, the coupling to the fundamental mode \cite{cohen1992atom}, $g_0$, scales with cavity length $L$ as $g_0 \sim 1/L$.
Holding constant the qubit frequency,
we are interested in the $nth$ harmonic of the cavity which is nearly resonant with the qubit. The ratio of the coupling for this mode to the FSR then scales as $g_n/FSR \sim \sqrt{L}$.
As there are many modes simultaneously interacting with the qubit, many modes can be coupled to each
other during the typical lifetime of an excitation in the system,
with the effective qubit mediated interaction rate between a pair of modes being
$\frac{g_mg_n}{(n-m)FSR}$,
leading to a delocalization of the
excitation across many modes.


\begin{figure}
\label{fig1}
\centering
\includegraphics{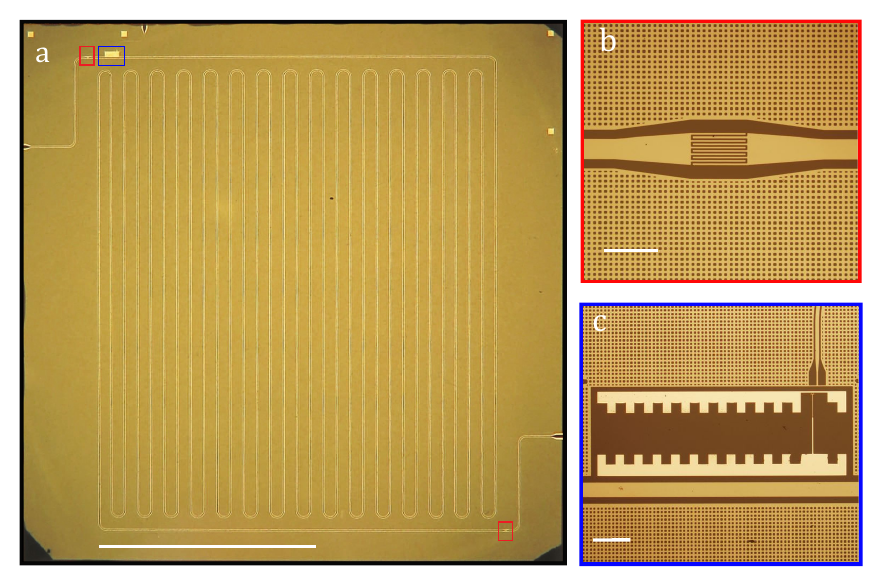}
\caption{\textbf{Device for achieving MMSC. }\textbf{a},  A low fundamental frequency, $\omega_0/2\pi $= 92 MHz, is realized by fabricating a 0.68 m coplanar waveguide resonator on a 25 mm x 25 mm sapphire substrate. Capacitive coupling at the input and output ports allows coupling of radiation into and out of the resonator, at rates $\kappa_m/2\pi$ $\sim$ 0.5-3 MHz. Red rectangles show coupling capacitors. Blue rectangle encloses qubit. \textbf{b}, Symmetric interdigitated  capacitors couple cavity to transmission lines.
\textbf{c}, A transmon qubit is capacitively coupled to the center pin near the output capacitor, an antinode for all modes.  SQUID geometry allows for tuning qubit frequency, $\omega_a$, via the flux bias line (shown) or the external magnet \cite{Koch2007} . The grid-like pattern in ground plane pins flux vortices. Scale bars denote (a) 10 mm, (b) 100 $\mu$m, and (c) 100 $\mu$m.}
\end{figure}

\begin{figure*}
\label{fig2}
\centering
\includegraphics[keepaspectratio=true, width=7in]{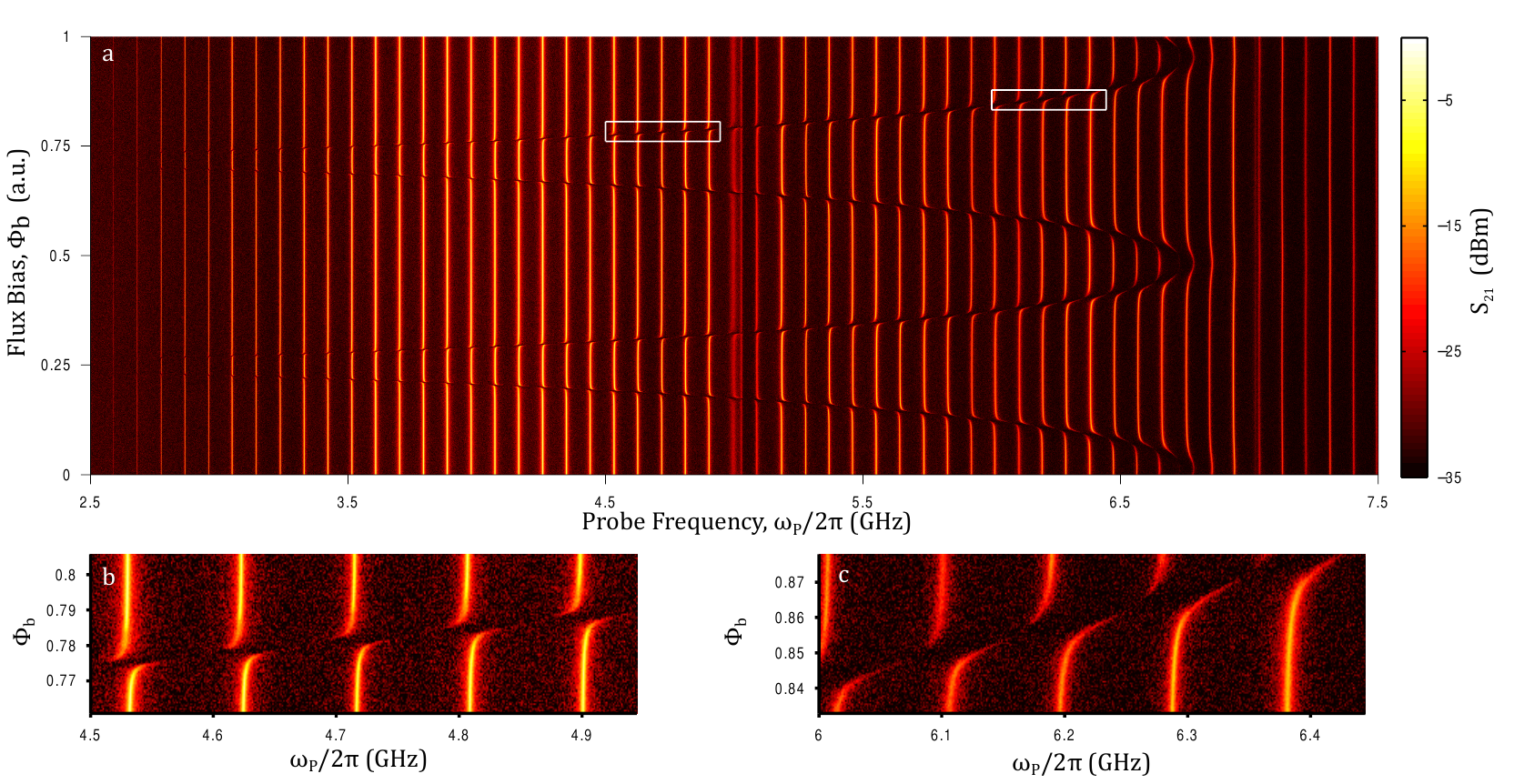}
\caption{\textbf{Demonstrating multimode strong coupling.} \textbf{a}, Cavity transmission as qubit frequency is tuned with an applied flux bias. Vacuum Rabi avoided crossings are visible as the qubit is tuned into resonance with each mode (bright vertical lines spaced by ~92 MHz).  It is apparent that the vacuum Rabi splitting is comparable with the free spectral range (mode spacing), thus indicating that MMSC has been achieved. White rectangles are shown in b, c. \textbf{b-c}, The coupling $g_m$ grows with m: $g_0\sqrt{m+1}$. A fit to the first manifold of the multimode Jaynes-Cummings yields $g_0 \sim 3.75$ MHz.}
\end{figure*}

Within this rich domain, we first focus our attention on resonance fluorescence, or the response of a qubit to coherent drive. Strongly driving a qubit in free space results in emission forming a three peaked structure known as the Mollow triplet, composed of a center peak and two symmetric
sidebands \cite{Mollow1969}. The sidebands are displaced linearly from the central peak by the Rabi
amplitude \cite{Savage1989} $\Omega$. This phenomenon, also observed experimentally with superconducting
qubits \cite{Astafiev2010}, is typically explained using the dressed state picture approach \cite{Cohen-Tannoudji1977}. Coupling of a qubit to a single mode cavity strongly modifies the Mollow
triplet \cite{Muller2007,Lang2011, Oelsner2013}. For weak coupling, the cavity passively filters the fluorescence, while at strong coupling, the sideband width is proportional to the coupling \cite{Savage1989,Kim2014}.

We design a very low frequency cavity such that the $\omega_a$ of a standard qubit falls near a high harmonic ($\sim$ 50th-75th) of the cavity, unlike traditional circuit QED experiments. Additionally, this approach allows us to adapt to the typical measurement range (3-8GHz) of existing circuit QED setups
(see Methods). The superconducting microwave cavity comprises a 0.68 m transmission line with identical capacitors on either end, resulting in a fundamental frequency, $\omega_0/2\pi = 92$MHz (see Fig.~1 and Methods section for details). We couple a flux-tunable transmon qubit \cite{Koch2007} near one end of the resonator, an anti-node for all the modes of the cavity, and achieve a qubit-mode coupling strength for the 75th cavity harmonic exceeding 30 MHz. Coherent drive, at frequency $\omega_d$, is introduced only via the input port (Fig.~1b), while radiative decay of cavity photons occurs via both input and output ports. A unitary transformation maps the coherent drive from the cavity to the qubit, yielding a qubit driving term
$\Omega \cos(\omega_d\,t)(\sigma^+ + \sigma^-)$, with $\Omega$ the Rabi drive amplitude (see Supplement).  The mode linewidths, $\kappa_m/2\pi$, span from 0.5-3 MHz to provide reasonable separation between modes. The qubit spontaneous decay rate, $\gamma/2\pi$, is of the order of 1.6 MHz, due to the multimode Purcell
effect \cite{Houck2008}.

Measuring transmission while tuning the qubit frequency reveals that the qubit strongly couples to a vast number of modes (Fig.~2a) \cite{Wallraff2004}. In fact, we are able to follow the trail of avoided crossings for several GHz as the qubit energy tunes through the uniformly dense mode structure. The smooth transitions between subsequent avoided crossings show that the qubit is perpetually near resonance with some mode and as such the dispersive approximation is never valid. In Figs. 2b-c, the marked difference in the magnitude of the avoided crossings, and thus coupling strength, is apparent. We apply a fit to find $g_0/2\pi \sim 3.75$ MHz, giving rise to $g_{75}/2\pi \sim$ 32 MHz. 

\begin{figure*}
\label{fig3}
\centering
\includegraphics{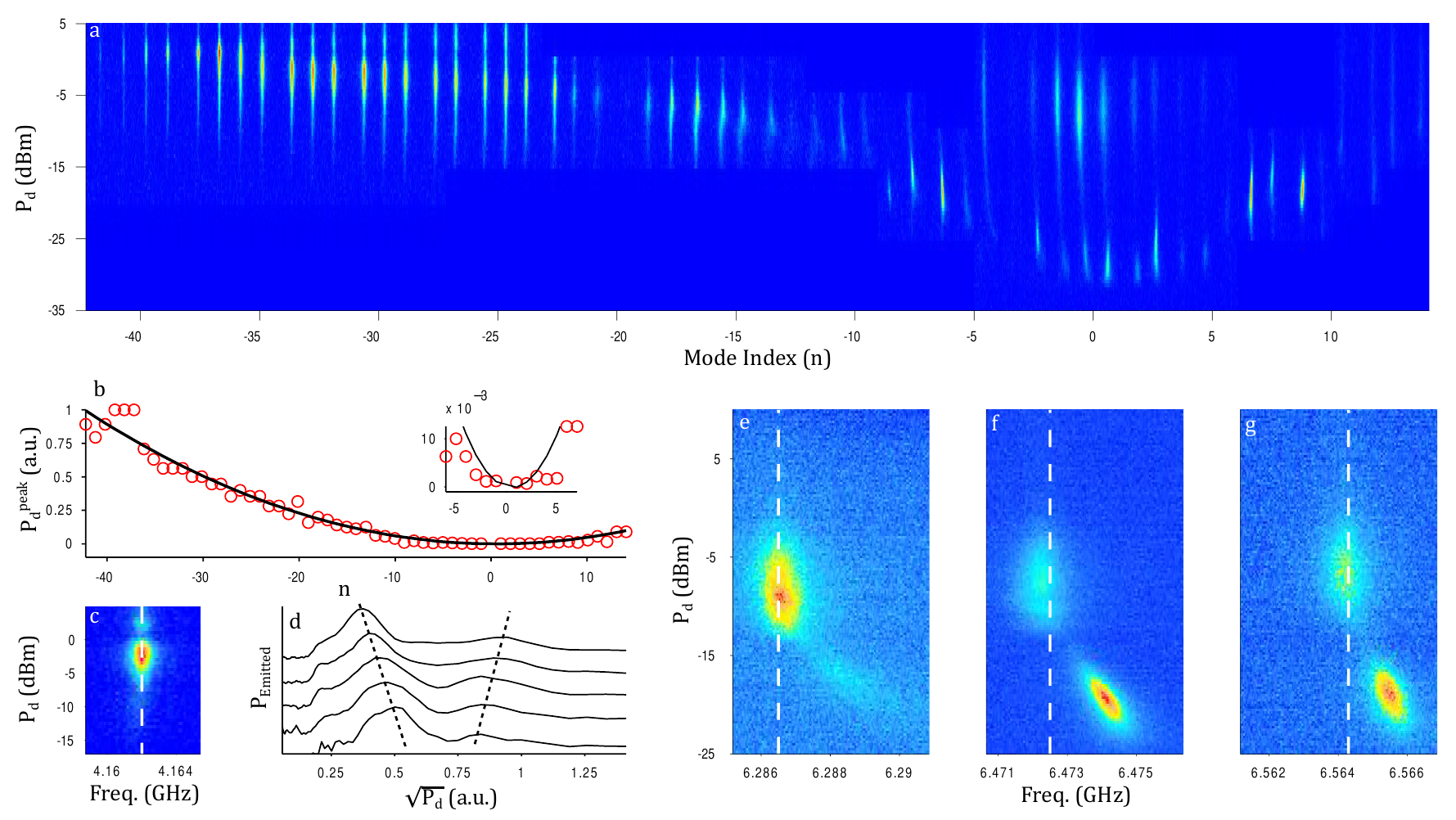}
\caption{\textbf{Multimode resonance fluorescence. }\textbf{a}, Power spectrum for varying drive power, $P_d$, when $\omega_{d} = \omega_a= \omega_{75}$, measured at more than 50 modes. Mode index $n$ is the relative detuning from the drive mode, $n= m -75$. Power is measured in a 10MHz window around each mode. Increasing drive power unveils fluorescence at farther detuned modes, strong emission from multiphoton processes, and complex multi-lobed fluorescence spectra. \textbf{b}, The drive power corresponding to peak fluorescence for each far detuned mode shows a quadratic dependence on detuning, suggesting this fluorescence is due to cavity enhanced sideband fluorescence \cite{Kim2014}. Modes near the drive frequency (see inset) do not follow the quadratic fit due to strong multimode interaction. \textbf{c}, Detailed plot of $n=-30$ from (a) shows a multi-lobed structure in fluorescence spectrum. Bare mode frequency is denoted with a vertical white dashed line. For \textbf{d}-\textbf{g}, $\omega_{d} = \omega_a= \omega_{69}$. \textbf{d}, We explore the difference between the two lobes seen in (c) for several modes ($n =  -36$, $-34$, $-32$, $-30$, $-28$, from bottom to top with individual curves vertically displaced for clarity) by measuring the total fluorescence power in a $6$ MHz window while varying drive power. 
The drive powers corresponding to peak fluorescence move oppositely with mode number for the two lobes.
\textbf{e-g}, For modes near the drive (n= -1, 1, 2), strong multimode interaction further yields a more complex multi-lobed structure. Bare cavity mode frequencies are denoted with vertical white dashed lines.}
\end{figure*}


In Fig.~3 we observe fluorescence across more than 50 modes,
when the drive and qubit are resonant with a high harmonic of the cavity.
Enhanced fluorescence is observed at mode $m$ when it is resonant with the Rabi sideband.
As the displacement of Mollow sidebands is proportional to drive amplitude $\Omega$, 
this occurs \cite{Kim2014} when detuning from the $m$th mode
$\Delta_{m} \equiv \omega_m - \omega_d \approx \Omega$.
For far detuned modes the drive power needed to reach the peak fluorescence at these modes
$P_d^{peak} \approx \Omega^2$, and hence $P_d^{peak} \approx \Delta_{m}^2 $ (Fig.~3b).


At fixed drive power, emission is spread over many modes,
with the simultaneous enhancement at many nearby modes, differentiating the multimode fluorescence from the single mode case.
With varying drive power, far detuned modes exhibit a multi-lobed fluorescence structure (e.g.~Fig.~3c).
The drive power for the peak of the secondary lobe approaches twice that of the first lobe for increasing detuning $\Delta_m$
(Fig.~3d), originating from multiphoton decays. Two photon decay processes in single mode cavities, where $\Omega = 2\Delta_{m}$, have previously been observed \cite{Lange1996, Ota2011}.
For MMSC, the multiphoton decay processes can progress via combinations of many different modes.
The resonance condition for single and multiphoton processes is thus shifted away from
$\Omega = p \; \Delta_{m}$, for $p \in \mathbb{Z}$ (see Supplement).

At smaller detuning, the drive power for the peak fluorescence deviates from a quadratic dependence on detuning.
Multimode effects become increasingly dominant as we approach modes closer to the qubit, with the appearance of a different intricate multi-lobed structure.
When the resonant mode is very weakly driven, there is no incoherent scattering to the other modes. However, at a higher drive power, fluorescence appears simultaneously at multiple nearby modes (three of which are displayed in Fig.~3e-g). In addition to detecting bright fluorescence at these nearby modes (three on either side of the center mode) we also observe incoherent scattering at the center mode itself.
Furthermore, there is a change in emission frequency at each of these modes, approaching the bare cavity mode from above with increasing drive power. This shift is qualitatively the same for modes on either side of the drive, and cannot be attributed to a simple AC Stark shift.

As drive power is further increased, the simultaneous fluorescence at the nearby modes fades away. At an even higher power, the emission at these modes reappears, but at the bare cavity mode frequencies. This power coincides with that at which the far detuned modes fluoresce, suggesting the
emergence of qubit mediated massively multimode correlation.

A remarkable observation of our experiment is the appearance of ultranarrow linewidths in emission.
The fluorescence at far detuned modes, captured in Fig.~3, have linewidths that are roughly equal to that of the bare modes. For modes nearest the qubit the fluorescence linewidth is smaller than $\kappa_m$, e.g.~for the nearest mode above being respectively 650 kHz vs. 1.1 MHz. Detuning the drive from the center mode unveils here a linewidth narrowing by over an order of magnitude to 65 kHz (Fig.~4a). The narrowest fluorescence occurs when the drive is detuned by $2$ MHz from the resonant mode. This unexpected behavior is robust to qubit frequency; the narrowed fluorescence produced with the $2$ MHz detuned drive remains even as the qubit frequency is tuned over $150$ MHz away from the cavity mode (see Supplement).

\begin{figure}
\centering
\includegraphics[keepaspectratio=true, width=3.45in]{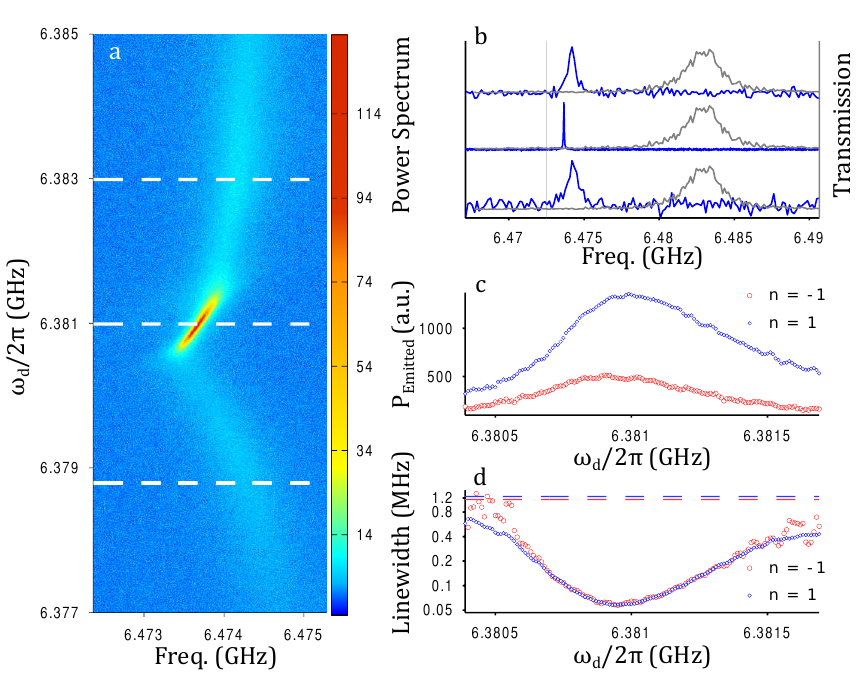}
\caption{\textbf{Ultranarrowing of fluorescence. }\textbf{a}, Power spectrum ($n=1$) for varying drive frequency ($\omega_d/2\pi$), at fixed drive power, reveals linewidth narrowing. Minimum linewidth, 65kHz, corresponds to $(\omega_d - \omega_{69})/2\pi = 2$ MHz. \textbf{b}, Normalized spectra along dashed lines in (a) are plotted in blue, vertically displaced. For comparison, low power transmission is plotted in solid gray and the bare mode frequency is denoted with a vertical gray line. \textbf{c}, Emitted power (over $\sim$ 3 MHz bandwidth) vs. $\omega_d/2\pi$ shows simultaneous brightening of $n=\pm1$ modes. \textbf{d}, Fluorescence linewidths vs. $\omega_d/2\pi$. Linewidths vary concurrently for $n=\pm1$ and are inversely proportional to emitted power. For comparison, dashed lines show $\kappa/2\pi$.}
\end{figure}

We attribute this narrowing to the spontaneous generation of coherence arising from the coupling of many dressed states through a common vacuum. Such behavior has been predicted and observed in atoms with multiple closely spaced bare atomic levels
coupled to vacuum or a single mode of a cavity \cite{Zhou1996,Kiffner2010,Heeg2013},
where the theoretical analysis is aided by the simpler dressed state structure.
In our setup, unlike the single mode case, the number of states in a single excitation manifold is not a constant but rather grows as
${N+M-1 \choose M-1} + {N+M-2 \choose M-1}$, where $M$ is the number of modes and $N$ is the excitation manifold. The rapid growth of the number of states per manifold hinders a simplified analysis via the dressed state picture, but makes possible the near resonance of many level spacings (Bohr frequencies), leading to a
collective enhancement of the coherences from their mutual couplings.
An external coherent drive allows us to dynamically access regions of the dressed states with
large generated coherences, as evident in the drive dependence of the narrowing in Fig.~4.

In the region where narrowing is observed, the linewidth and steady state photon number are observed to be inversely proportional, as has been previously suggested for the case of a single mode \cite{Freedhoff1994}. For our multimode cavity, the narrowing is in fact observed simultaneously for multiple modes at the same drive frequency, as depicted in Figs. 4c-d for the two modes nearest the drive.
While the modes directly neighboring the center mode exhibit the
sharpest linewidths of $65$ kHz, the next nearest modes also display a narrowed fluorescence, $\sim 300$ kHz. The multi-lobed power dependent structure of fluorescence also appears at the optimal detuned drive frequency, however, the linewidth of the second lobe is of order $\kappa_m$, for the nearby modes. For more distant modes, while fluorescence follows a similar $\Delta_{m}^2$ dependence, there is no narrowing apparent. Finally, the drive dependent narrowing is not unique to these described experimental parameters. When we translate the drive and qubit to another set of modes, we observe equally narrow linewidth fluorescence which indicates that spontaneous generation of coherence is generic in this system.

In this letter, we realized a new regime of circuit QED,
multimode strong coupling.
We demonstrated the creation of many-body steady states with long lived quantum coherence, as evidenced
by multimode resonance fluorescence with ultranarrow linewidths.
Dynamically generated quantum correlations between modes in this driven dissipative system
is expected to lead to qualitatively new physics \cite{Verstraete2009}.
While this new complex regime poses many theoretical challenges, precise circuit QED experiments can guide this exploration. We hope that access to multimode strong coupling will lead to many new advancements in circuit QED and our understanding of light-matter interactions.


\centering
\vspace{2mm}
\textbf{\\Methods\\}
\vspace{2mm}
\justify

The MMSC device was made using a combination of standard optical and electron-beam lithography techniques on a 25 mm x 25 mm sapphire wafer. The cavity and larger features were etched from a 200 nm thin film of sputtered Niobium, while the qubit was made with evaporated Aluminum using the Dolan Bridge technique. The cavity has a center pin to gap ratio of 10:4.186. The transmon qubit \cite{Koch2007} is capacitively coupled near the end of the cavity, at an antinode. The qubit consists of two large islands and a pair of parallel junctions in a SQUID-loop geometry. While there is an on-chip flux bias line, a superconducting magnet was installed around the device to tune the qubit frequency without detrimental thermal heating. The device was extensively wire-bonded to the circuit board to ensure proper connection of all ground planes and reduce the effect of spurious modes. The device was cooled to ~11 mK in a dilution refrigerator, and measured using standard low-temperature microwave circuit elements including an IQ mixer, isolators, and a HEMT amplifier.\\

\centering
\vspace{2mm}
\textbf{{Acknowledgements\\}}
\vspace{2mm}
\justify

We thank C. Eichler, S. M. Girvin, J. Koch, D. I. Schuster, W. E. Shanks, S. J. Srinivasan, and J. M. Taylor for helpful conversations and technical support.

\centering
\vspace{2mm}
\textbf{{\\Funding\\}}
\vspace{2mm}
\justify

This work was supported by 
the US Army Research Office (W911NF-11-1-0086), 
the US National Science Foundation through the Princeton Center for Complex Materials (DMR-0819860) and CAREER awards (Grant Nos. DMR-0953475 \& DMR-1151810),
the David and Lucile Packard Foundation,
and the DoD through the NDSEG program.
There are no competing financial interests.

\centering
\vspace{2mm}
\textbf{{\\Author contributions\\}}
\vspace{2mm}
\justify

N. M. S. fabricated the device, led the experiments, analyzed the data, and wrote the manuscript. Y. L. and D. S. provided key contributions in the experiments and data analysis, helped in writing the manuscript, and performed the numerical simulations. L. J. S. contributed to cavity design. D. U. provided technical advice and built the superconducting magnet. M. M. and H.E.T. provided theoretical support. A. A. H. advised and supervised the project.



\renewcommand{\figurename}{Figure S\!}

\newpage
\centering
\vspace{2mm}
\textbf{{SUPPLEMENTARY MATERIAL}}
\justify

\ifpdf
\DeclareGraphicsExtensions{.pdf, .jpg, .tif}
\else
\DeclareGraphicsExtensions{.eps, .jpg}
\fi


\tableofcontents

\section{Experimental Details}
\subsection{Fridge Diagram}

The experimental measurement setup is shown in Fig. S1. 

\subsection{Standard Spectroscopy}
Two-tone spectroscopy or dispersive readout is a standard measurement technique which uses the dispersive shift of a mode to precisely determine the frequency of the qubit~\cite{Bishop2008}. 
In our massively multimode strong coupling experiment, such spectroscopy shows that
the mode frequency is affected not only by the qubit state, but also by populating the modes nearby to the qubit. This is yet another manifestation of mode-mode interactions mediated via the qubit. The magnitude of transmission, a proxy for the dispersive shift, is indicated by the colorbar in Fig. S2.
As the qubit frequency approaches a mode, the strength of interaction between that mode and the measurement mode steadily increases.

This interaction is so strong that the measurement mode shifts substantially even when the qubit is over 200 MHz away. Thus the frequency of the measurement mode needed to be recalibrated for each flux voltage in Fig. S2. Furthermore, to track the qubit in this frequency range with sufficient contrast, the designated measurement mode needed to be changed such that it was farther and farther away from the qubit. For each change of measurement mode, there is a horizontal black line in Fig. S2.

\subsection{Ultranarrow Linewidth Versus Qubit and Drive Frequencies}

We find that ultranarrow fluorescence is very robust to the flux bias voltage of the transmon qubit as shown in Fig. S3, where we display the linewidth as a function of the flux bias voltage applied to the qubit.
This observed robustness will be discussed in the following theoretical analysis.
Also displayed is the integrated fluorescence power against the same bias. 
The power and linewidth are seen to be inversely related.
The drive frequency and power here are both fixed to that which results in the minimum linewidth in Fig. 4 of the main text.

\section{Theoretical Discussion}
\subsection{Effective Hamiltonian and Master Equation}
In order to better interpret the intriguing dynamical phenomena observed in our experiment, here we provide a simplified model to illustrate key aspects of the physics. Even though the model is far from a complete representation of the experimental system, we argue that qualitatively similar physics emerges in the massively multimode system. 

This driven dissipative system can be described by the following master equation ($\hbar=1$ in our units)
\begin{widetext}
\begin{align}
\frac{\partial\rho}{\partial t} &=
-i \left [H, \rho \right] + \frac{\gamma}{2}
\left( 2\sigma^{-}\rho\sigma^{+} - \sigma^{+}\sigma^{-}\rho - \rho\sigma^{+}\sigma{-} \right) 
\:+\: \sum_{m} \frac{\kappa_m}{2}
\left( 2a_m\rho a^{\dagger}_m - a^{\dagger}_ma_m\rho - \rho a^{\dagger}_ma_m \right) \ , \\
H &= H_0 +  i\eta \left( a^{\dagger}_{r} - a_{r} \right) \ ,\\
H_0 &= \sum_{m} \: \Delta_m \: a_{m}^{\dagger}a_{m} + \Delta_a \: \sigma^{z} +
g_m \left( a^{\dagger}_{m}\sigma^{-}  + \text{h.c.} \right) \ ,
\end{align}
\end{widetext}
with $\kappa_m$ the photon loss rate for mode $m$, and $\gamma=1/T_1$ the qubit
spontaneous decay rate.
The above Hamiltonian is written in the rotating frame of the drive, and the rotating wave approximation (RWA) has been used.
$\Delta_m = \omega_m - \omega_d$ is the detuning between the $m$th mode and the drive. Likewise,
$\Delta_a = \omega_a - \omega_d$ is the detuning between the qubit and the drive. RWA is a good approximation if we focus on high harmonics ($\sim70$ in the experiment) of this multimode system. 
$\eta$ is the amplitude of the external drive, and the drive is assumed to be resonant with the $r$th mode. Steady state photon numbers of other modes in the uncoupled system ($g_m=0$), are suppressed relative to the resonant term, by multiples of $(FSR/\kappa)^2$, and we are justified in keeping only the resonant
drive.
Applying an active drive and dissipation dependent unitary displacement
to the state of the system, via
$\tilde{\rho} = D^\dagger\rho D$, with
\begin{eqnarray}
D = \exp \left(\xi (a^{\dagger}_r - a_r) \right) \ , \ \ \ \xi = \frac{2\eta}{\kappa_r} \ ,
\end{eqnarray}
transfers the drive in the master equation onto the qubit. The Hamiltonian appearing in
the master equation is now $H = H_0 + \Omega(\sigma^{+} + \sigma^{-})/2$, where $\Omega = 8g_r\eta/\kappa_r$ is the Rabi amplitude. We could in fact directly implement such a drive in our experiment by pumping the flux bias line to qubit. 
For the sake of simplicity we start from this form of the Hamiltonian (externally driving the qubit)
in what follows, and study a restricted two mode system, also taking the drive to be resonant with the
qubit. We disregard the mode resonant with the qubit, focusing on the more interesting sideband phenomena.

The master equation for the simplified two mode model is
\begin{widetext}
\begin{align}\label{eq:master equation}
\frac{\partial\rho}{\partial t} &= -i \left[H, \rho\right] + \frac{\gamma}{2}
\left(2\sigma^{-}\rho\sigma^{+} - \sigma^{+}\sigma^{-}\rho - \rho\sigma^{+}\sigma{-}\right)
\:+\: \sum_{m=1,2}  \frac{\kappa_m}{2}
\left(2a_m\rho a^{\dagger}_m - a^{\dagger}_ma_m\rho - \rho a^{\dagger}_ma_m \right) \ , \\
H &= \Delta_1 \: a_{1}^{\dagger}a_{1} + \Delta_2 \: a_{2}^{\dagger}a_{2} + \frac{\Omega}{2}
\left(\sigma^{+}+\sigma^{-}\right) + g_1\left(a^{\dagger}_{1}\sigma^{-} + \text{h.c.}\right) +
g_2 \left(a^{\dagger}_{2}\sigma^{-} + \text{h.c.} \right) \ .
\end{align}
\end{widetext}
Without cavity modes, the qubit dynamics result in the famous Mollow triplet. 
While the single mode case, for a special choice of parameters,
can be solved analytically ($\Omega=\Delta$)~\cite{PRA1993}, this is in general not possible for a general multimode system. 
We thus resort to numerical simulation based on the above equation \eqref{eq:master equation} in the following discussion.

To motivate the numerical exploration of the two mode master equation, we first analyze some
properties of this Hamiltonian to gain intuition.
Introducing a new basis for the qubit by defining
$\ket{\tilde{0}} = \frac{1}{\sqrt{2}}\left(\ket{0} - \ket{1}\right),
\ket{\tilde{1}} = \frac{1}{\sqrt{2}}\left(\ket{0} + \ket{1}\right)$,
yields the Hamiltonian
\begin{widetext}
\begin{eqnarray}
H =
\frac{\Omega}{2}\tilde{\sigma^{z}} +
\sum_{i=1}^2
\Delta_i \: a_i^\dagger a_i +
\frac{g_i}{2} \Bigg[\left(\tilde{\sigma}^{z} - \tilde{\sigma}^{+} + \tilde{\sigma}^{-}\right)a_{i}^{\dagger}
+ \text{h.c.}\Bigg] \ .
\end{eqnarray}
\end{widetext}
Applying an atomic-state-dependent mode displacement operator
$U = \prod_{i=1}^2 \exp \left(\frac{g_i}{2\Delta_i}(a_{i} - a_{i}^{\dagger})\tilde{\sigma_{z}} \right)$
to remove $\tilde{\sigma^{z}}$ from the interaction, we have
\begin{widetext}
\begin{eqnarray}\label{eq:interaction}
H = \frac{\Omega}{2}\tilde{\sigma^{z}} +
\sum_{i=1}^2
\Delta_i \: a_i^\dagger a_i +
\frac{g_i}{2} \Bigg[ \left(\tilde{\sigma^{+}}V_1 V_2 - \tilde{\sigma^{-}}V_1^{\dagger}V_2^{\dagger}\right)\left(a_i-a^{\dagger}_i\right) \Bigg] \ ,
\end{eqnarray}
\end{widetext}
where
\begin{align}
V_i &= \exp\left(\frac{g_i}{\Delta_i}(a^{\dagger}_i - a_i)\right) \\
&= \exp\left(-(g_i/\Delta_i)^2/2\right)\sum_{m,n}\left(g_i/\Delta_i\right)^{m+n}\frac{(-1)^n}{m!n!}a^{\dagger m}_ia^{n}_i .
\end{align} 
This form of the Hamiltonian reveals the complicated multimode multiphoton interactions of arbitrary order. 

To further simplify the analysis, as in the RWA, we truncate the Taylor expansion to remove fast oscillating terms. These approximations are only valid in the limit of $g_i/\Delta_i \ll 1$, and near certain resonance conditions. We thus arrive at a finite order multimode multiphoton Hamiltonian.
For example, let us assume that $\Delta_1 = -\Delta_2 = \Delta, g_1 = g_2 = g, \kappa_1 = \kappa_2 = \kappa$, and that the qubit is resonant with the drive frequency. In this special case, the resonance conditions are simply $\Omega \approx m\Delta$, for $m \in (1,2)$.
The effective Hamiltonians for the two different Rabi rates ($\Omega=\Delta$ and $\Omega=2\Delta$) become
\begin{widetext}
\begin{align}
\label{single-photon-ham}
H_{eff}^{(\Omega=\Delta)} &=  \Delta a_{1}^{\dagger}a_{1} - \Delta a_{2}^{\dagger}a_{2} + \frac{\Delta}{2}\tilde{\sigma^{z}} + \frac{g}{2}[\tilde{\sigma}^{+}a_1 - \tilde{\sigma}^{+}a^{\dagger}_2+ \text{h.c.}] \ , &m=1\\
\label{two-photon-ham}
H_{eff}^{(\Omega=2\Delta)} &= \Delta a_{1}^{\dagger}a_{1} - \Delta a_{2}^{\dagger}a_{2} + \Delta\tilde{\sigma^{z}} + \frac{g^2}{2\Delta}[-\tilde{\sigma}^{+}a_1^2 + \tilde{\sigma}^{+}{a^{\dagger}_2}^2+ \text{h.c.}] \ . &m=2
\end{align}
\end{widetext}
Note that two photon processes in the second Hamiltonian
\eqref{two-photon-ham} above are supressed relative to the single photon ones in \eqref{single-photon-ham}
by $g/\Delta$.
Terms like $ \tilde{\sigma}^{+}a^{\dagger}_2 $ arise because the second mode carries negative frequency in the rotating frame.
The second Hamiltonian is the basis for two photon lasing of driven atoms in a cavity~\cite{PRL1990}. In our case where only a single qubit is present, keeping high-order quantum correlations beyond semiclassics is essential in explaining relevant coherence properties of the fluorescence light. The full master equation could be derived by applying these transformations on the dissipative part, giving rise to complicated multimode multiphoton dissipative terms.
We instead jump directly to the numerical results derived from the two mode master equation
\eqref{eq:master equation}.

\subsection{Multimode Multiphoton Processes}
We use the Monte Carlo Wavefunction approach to numerically simulate the dynamics of this driven dissipative multimode system.
In our simulation, the size of the Hilbert space needed grows as $2\prod_{m} N_m$,
with $N_m$ the photon cutoff for mode $m$. The rapid growth of the dimension of the Hilbert limits the number of modes that can be incorporated into the simulation. Expectation values of relevant operators are followed and steady state values for the driven system are presented below. To compute two time correlation functions such as $\langle a^{\dagger}_{1,2}(\tau) \: a_{1,2}(0) \rangle$ using this method, we have to employ a two step procedure~\cite{RMP1998}, equivalent to the quantum regression theorem, but at the expense of requiring large numbers of walks to produce reliable results. 

To explain the lobe structure and multimode fluorescence observed in our experiment, we first analyze two special limits of the two mode model. Resonant enhancement by multiphoton processes in a single mode $i$
(via the interaction $\sigma^{+}a^{m}_i + \text{h.c.})$, while tuning the qubit drive $\Omega$, leads to peaks in the steady state photon number near subharmonics of the Rabi rate $|\Delta_i| = \Omega/m, m \in \mathbb{Z}$. 
This is the physical origin of the lobe structure at far detuned modes in our massively multimode system, as $g/\Delta \ll 1$ is a good approximation for these modes.
Fig.~S4 presents numerical results for the special two mode case where the two modes are situated symmetrically around the qubit (and the drive which is resonant with it). Fluorescence at these two modes are equally and simultaneously enhanced. Moreover, the two peaks in photon number and qubit polarization
$\sigma^+$ deviate from the bare resonance conditions $|\Delta| = \Omega/m, m = 1, 2$. The qubit inversion
$\sigma^z$ can go above $0$. The results for a single mode model are also shown for comparison. While multiphoton resonances are apparent in both cases, the two mode system reaches much larger photon numbers in the steady state. The difference also appears in the spectral linewidth (see Fig.~S6). While the single mode fluorescence shows no spectral narrowing, the spectral linewidth in the two mode model can be much smaller than the natural linewidth. 
Beside the simple two photon interaction in a single mode ($\sigma^{-}a^{\dagger}_ia^{\dagger}_i$) analyzed above, the interaction vertex coupling two different modes ($\sigma^{-}a_ia^{\dagger}_j, i\not=j$) will result in resonant enhancement in modes near harmonics of the Rabi rate $|\Delta| = m\Omega$. This is demonstrated in Fig.~\ref{fig:Photon Number_II} where another special two mode case ($\Delta_2 = 2\Delta_1$) is investigated. Similarly to the previous case, this effect leads to co-enhancement of fluorescence at two modes, as evidenced by the brightening of the second mode at $\Delta \sim 80~\mathrm{MHz}$. As this is a second-order effect induced by the first mode, $N_2$ is much smaller than $N_1$. For stronger coupling, we would expect the difference to be smaller. 

In our strong coupling multimode system, all multimode multiphoton terms have to be taken into account, which makes the analysis very cumbersome. The competition between these processes will generally make resonances broader and less prominent. The numerical simulations qualitatively agree with our experimental results. 

\subsection{Multimode Spectral Narrowing}
Ultranarrow resonance fluorescence has been predicted for the single mode cavity \cite{PRL1994}, where coherence between dressed states of the driven system arises from coupling to common vacuum fluctuations. The resultant spectrum linewidth of the mode could be much smaller than the natural linewidth of the bare system. In fact, the linewidth of the sideband is proportional to the inverse of the steady state photon number $1/N$. We provide numerical evidence that this is also true in the multimode system. It is worth noting that reference~\cite{PRL1994} only considered a special value of the parameters, where $\Omega = \Delta$ and $g \sim \gamma$. Here we generalize their result to explain our observations. 

In Fig.~S6, for the symmetric two mode system, it is shown that spectral narrowing appears for a range of the Rabi rate $\Omega$. Moreover, the linewidth is inversely proportional to the photon number (see also Fig.~S4(a)) such that the linewidth reaches its minimum as the photon number reaches its maximum at $\Omega = 106~\mathrm{MHz}$. The left mode also shows the same amount of spectral narrowing (data not shown). 
The spectral narrowing in this simplified model doesn't reach that of our experiments (a factor of $\sim 3.3$ in the simulation vs. $\sim 17$ in the experiment), yet it displays qualitively similar features. Moreover, the relation $\kappa_{fluo} \sim 1/N$ suggests that the linewidth could be made much narrower than that of the current simplified model, when reaching larger steady state photon numbers.
For the interacting massively multimode system, the proliferation of the dressed states in the high excitation manifolds leads to enhanced generation of coherence among the dressed states
from a collective interference, which relies on the near resonance of many dressed state level
spacings. Concomitant with the collective suppression of dissipation for the steady state then is the rise in the steady state occupation.
Simulations of such a massively multimode system demands more computational resources due to the growth of the Hilbert space. More extensive results will be presented elsewhere.

Spectral narrowing is generally very insensitive to the qubit frequency. For finite drive-qubit detuning $\Delta_a$, the effective Rabi rate is $\Omega = (\Omega^2_0 + \Delta_a^2)^{1/2}$, with $\Omega_0$ the
amplitude of the drive. Based on the above analysis, spectral narrowing occurs when $\Omega$ is within the range $\Delta \pm \delta(\Omega_0)$, where
$\Delta$ is again the mode frequency measured with respect to the drive (i.e. its detuning),
and $\delta(\Omega_0)$ is generally determined by the number of modes, the coupling strength and the dissipation.
For relatively small detuning $\Delta_a$ and large Rabi rate $\Omega_0$, this gives a large effective range of qubit frequency $\delta(\Delta_a) \sim \frac{\Omega}{\Delta_a}\delta(\Omega_0)$ over which narrowing
occurs. 
It is worth noting that even if the drive is not resonant with the qubit and the two modes are not symmetric around the drive, spectral narrowing still exists in our simulation.

For a single mode system, it was argued that spectral narrowing requires that the qubit decay rate $\gamma$ be comparable to the coupling strength $g$, i.e. the moderate coupling regime~\cite{PRL1994}. We therefore study the effect of qubit decay rate $\gamma$ on steady state properties of the two mode system. Fig.~S7(a) presents photon number $N_1$ versus Rabi rate for a different $\gamma$ value. From this result, we conclude that spectral narrowing survives when going from moderate coupling to the strong coupling regime,
which we attribute to the fact that the eigenstates of the multimode system in high excitation manifolds are more densely spaced than in the single mode system. Dynamically generated coherence between these closely spaced states then results in spectral narrowing of many modes simultaneously.

\setcounter{figure}{0}

\begin{figure*}
\centering
\includegraphics[scale=0.8]{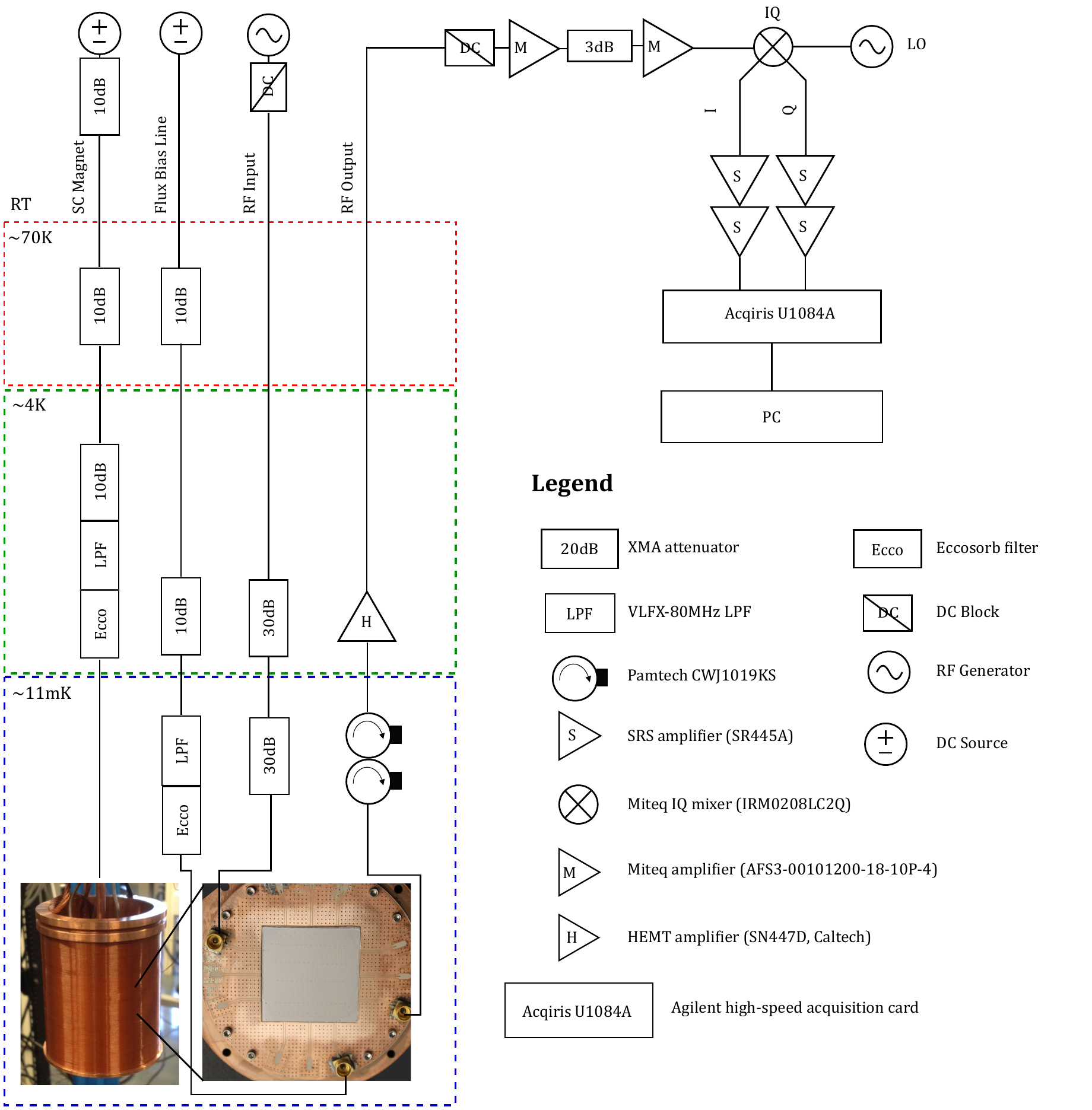}
\caption{Schematic representation of measurement setup. Device is mounted within the superconducting magnet, inside a dilution refrigerator. Experimental results presented in the paper were obtained from this general setup, with varied digital processing.}
\end{figure*} 

\begin{figure*}
\centering
\includegraphics[scale=1]{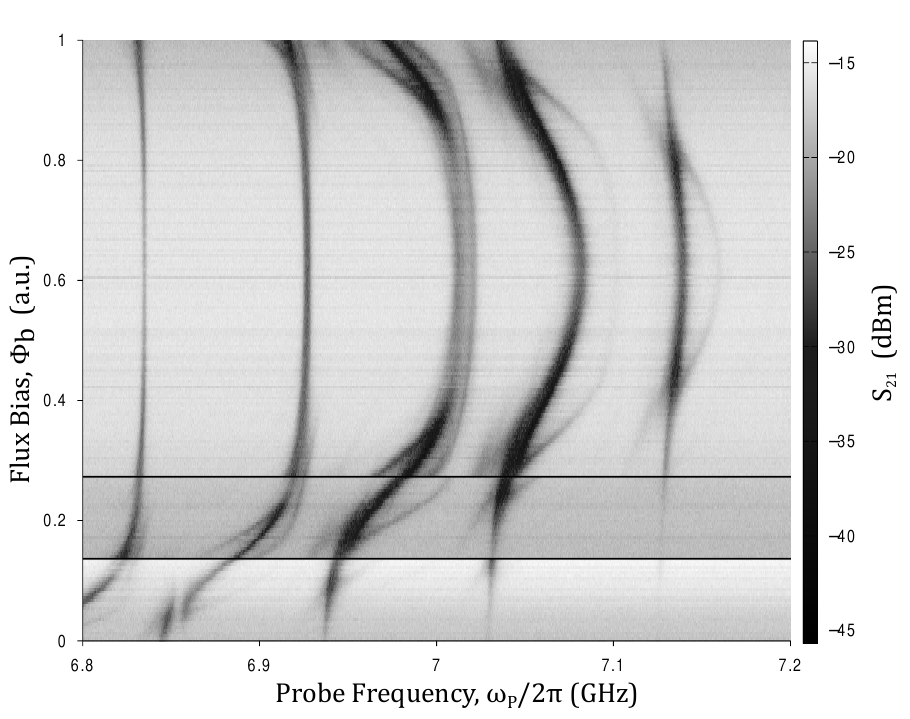}
\caption{Mode-mode interaction in dispersive readout. Tracking the change in dispersive shift for measurement mode, at $\omega_M$, with varying probe frequency, $\omega_p$, shows the impact of populating nearby modes. The magnitude of the dispersive shift is related to the detuning of a mode from the qubit (at 7.08GHz for $\phi_b = 0.6$). The qubit frequency is tuned using the on-chip flux bias line. Fainter curves are attributed to higher order processes. Horizontal black lines indicate a change of $\omega_M$ (6.469, 6.653, and 6.747 GHz from top to bottom), chosen to maintain sufficient distance from the qubit.}
\end{figure*} 

\begin{figure*}
\centering
\includegraphics[scale=1.0]{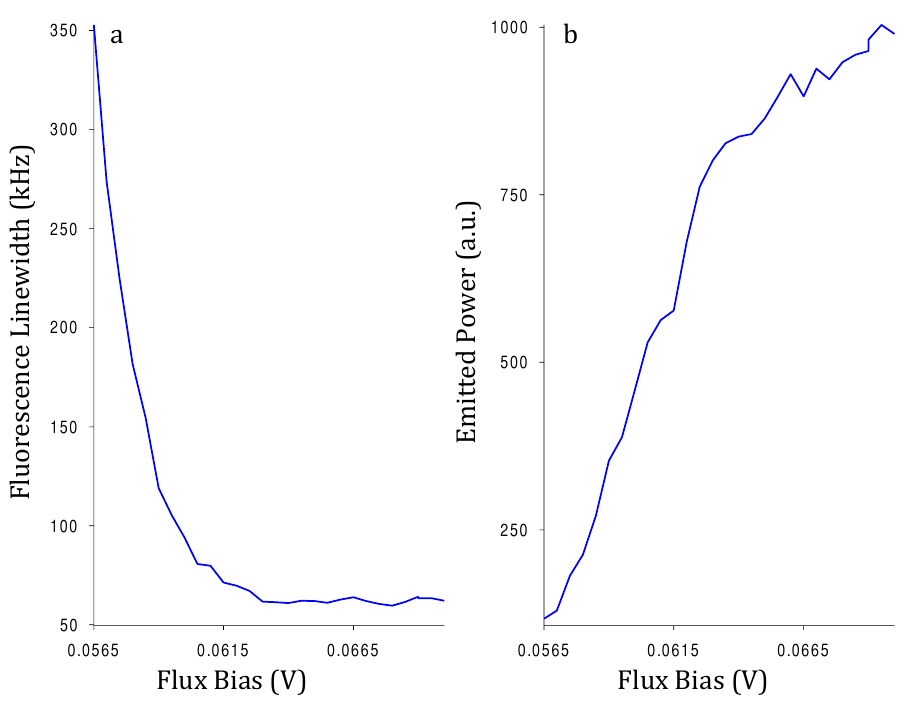}
\caption{Measured linewidth and emitted power versus flux bias. Spectral narrowing appears over a wide range of flux bias, corresponding to over $150~\mathrm{MHz}$ of the qubit frequency. The other parameters are the same as in Fig. 4 in the main text.
The fluorescence power and the linewidth are inversely related.}
{\label{fig:measured linewidth}}
\end{figure*} 

\begin{figure*}
\centering
\includegraphics[scale=0.6]{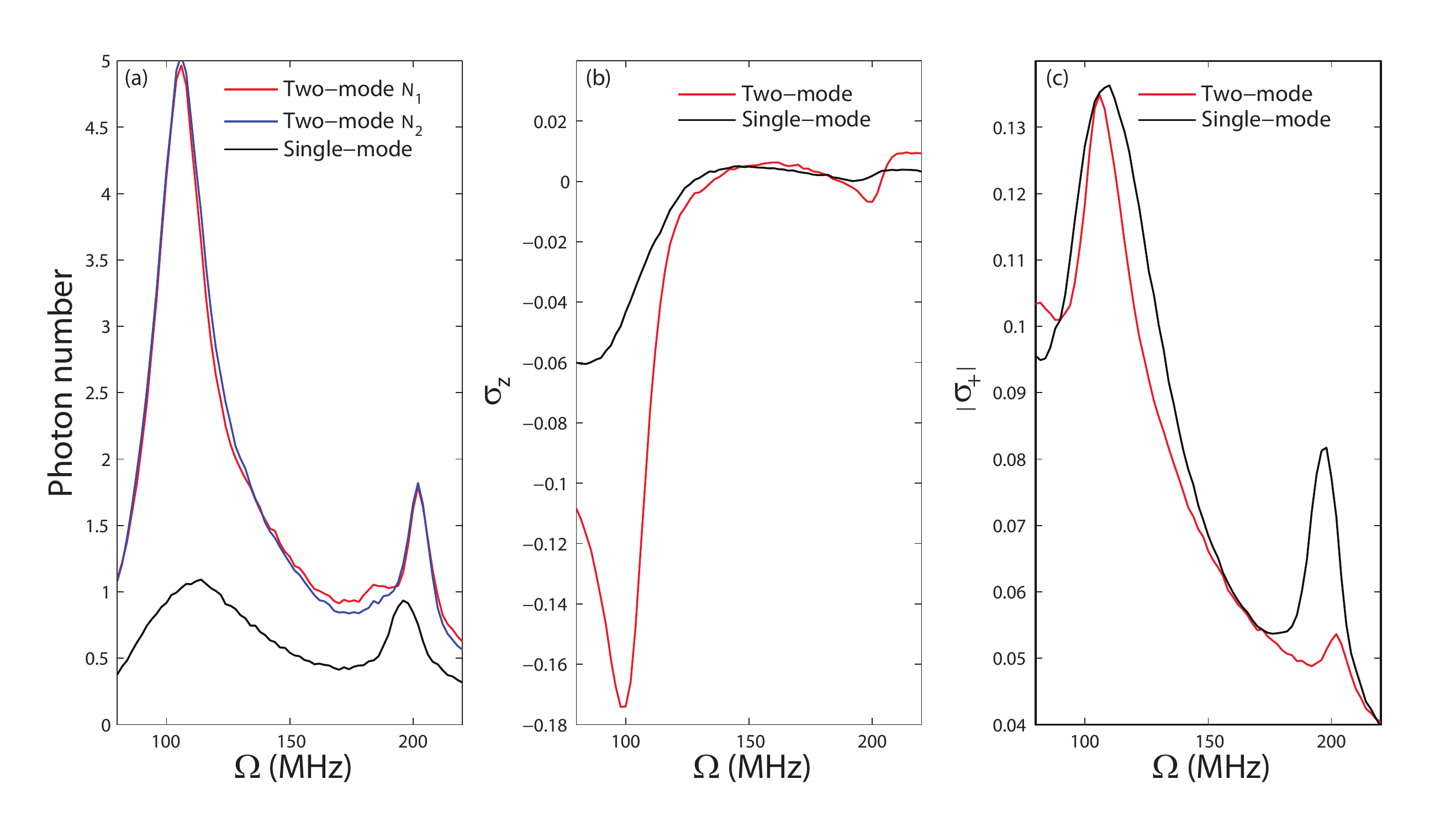}
\caption{Steady state photon number, qubit inversion and polarization, first parameter set. Parameters of the two mode system are $\Delta_1/2\pi = -\Delta_2/2\pi = 100~\mathrm{MHz}, \kappa_1/2\pi = \kappa_2/2\pi = 1~\mathrm{MHz}, \gamma/2\pi = 15~\mathrm{MHz}, g_1/2\pi = g_2/2\pi = 15~\mathrm{MHz}$. Parameters of the single mode system are $\Delta/2\pi = 100~\mathrm{MHz}, \kappa/2\pi = 1~\mathrm{MHz}, \gamma/2\pi = 15~\mathrm{MHz}, g/2\pi = 15~\mathrm{MHz}$.
These plots include 24 different Rabi rates $\Omega$ with 1,000 quantum walks per Rabi rate in the numerical simulations.}
{\label{fig:Photon Number}}
\end{figure*} 

\begin{figure*}
\centering
\includegraphics[scale=0.6]{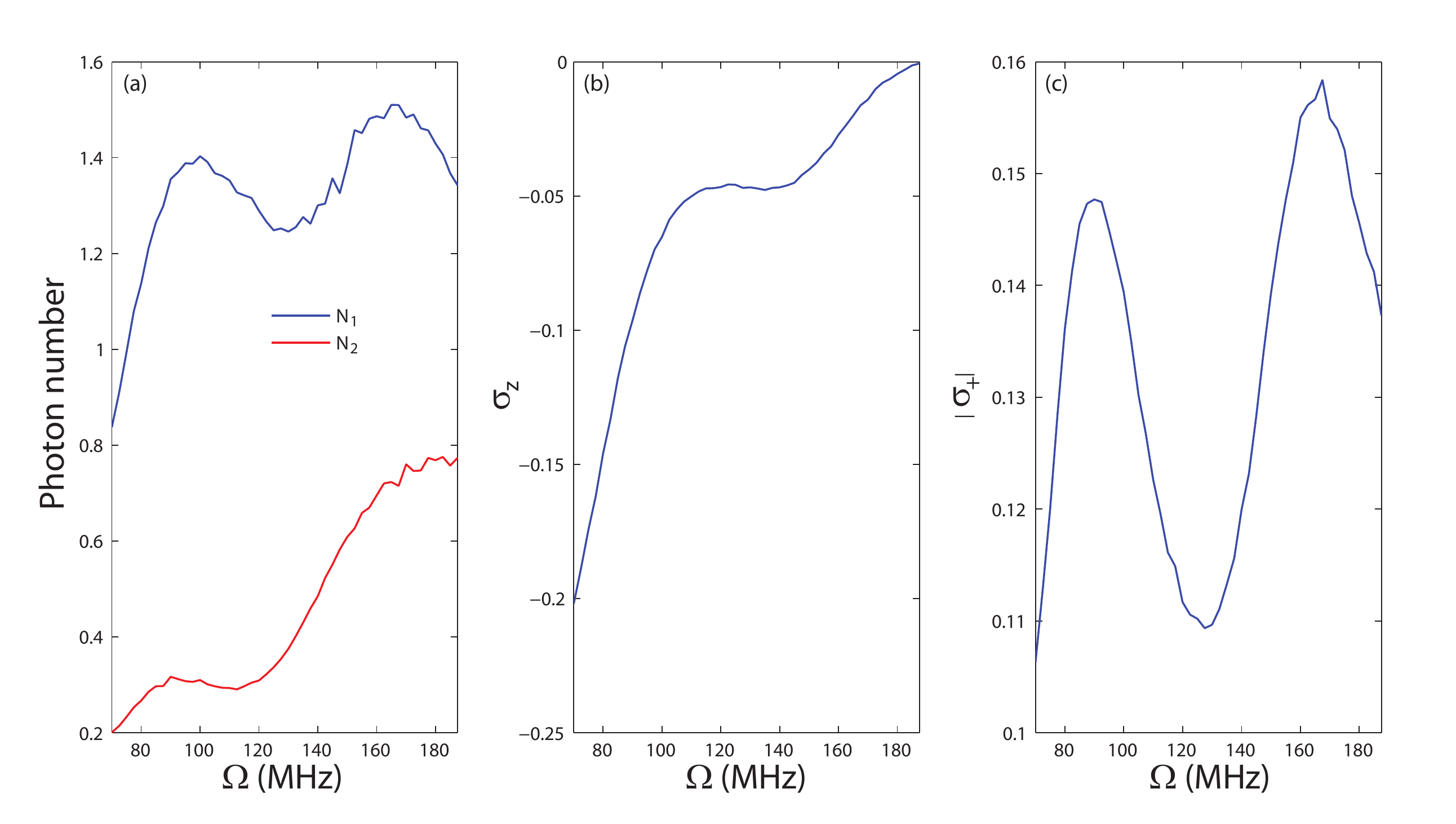}
\caption{Steady state photon number, qubit inversion and polarization, second parameter set. Parameters of the two mode system are $2\Delta_1/2\pi = \Delta_2/2\pi = 160~\mathrm{MHz}, \kappa_1/2\pi = \kappa_2/2\pi = 1~\mathrm{MHz}, \gamma/2\pi = 15~\mathrm{MHz}, g_1/2\pi = g_2/2\pi = 25~\mathrm{MHz}$. These plots include 24 different Rabi rates $\Omega$ with 1,000 quantum walks per Rabi rate in the numerical simulations.}
{\label{fig:Photon Number_II}}
\end{figure*} 

\begin{figure*}
\centering
\includegraphics[scale=0.6]{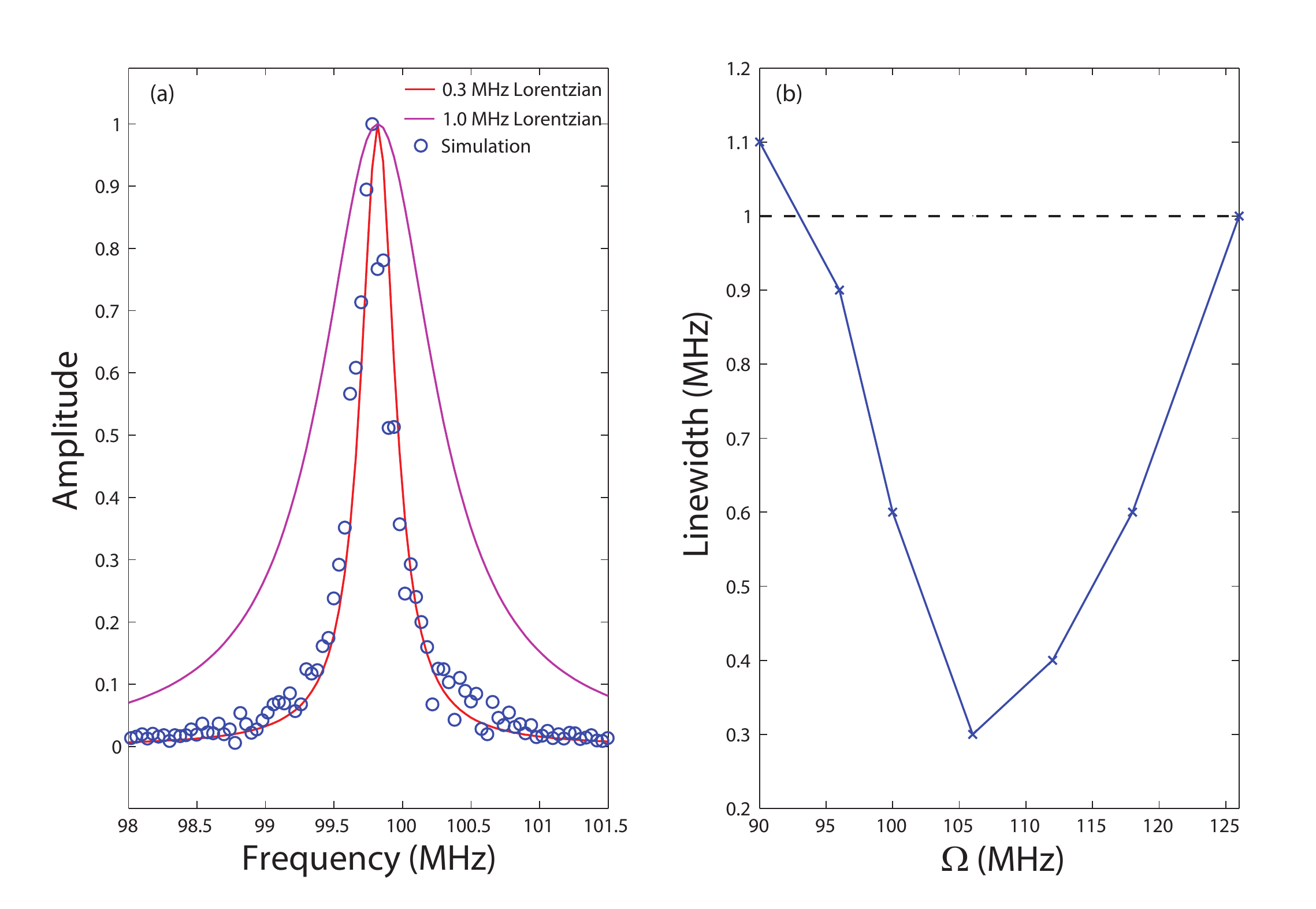}
\caption{Fluorescence spectrum of the right mode in a two mode system. (a) Parameters of the simulation are the same as Fig.~S4, with $\Omega/2\pi = 106~\mathrm{MHz}$, where the photon number reaches its peak. This figure includes 12,000 quantum walks in the numerical simulation. The spectrum is generated by Fourier transforming the two time correlation function of the mode operator. The fluorescence linewidth is more than $3$ times narrower than the natural mode linewidth. (b) Fluorescence linewidth versus Rabi rate. The black dashed line indicates the natural linewidth of the mode ($1~\mathrm{MHz}$).}
{\label{fig:Spectrum Narrowing}}
\end{figure*} 

\begin{figure*}
\centering
\includegraphics[scale=0.7]{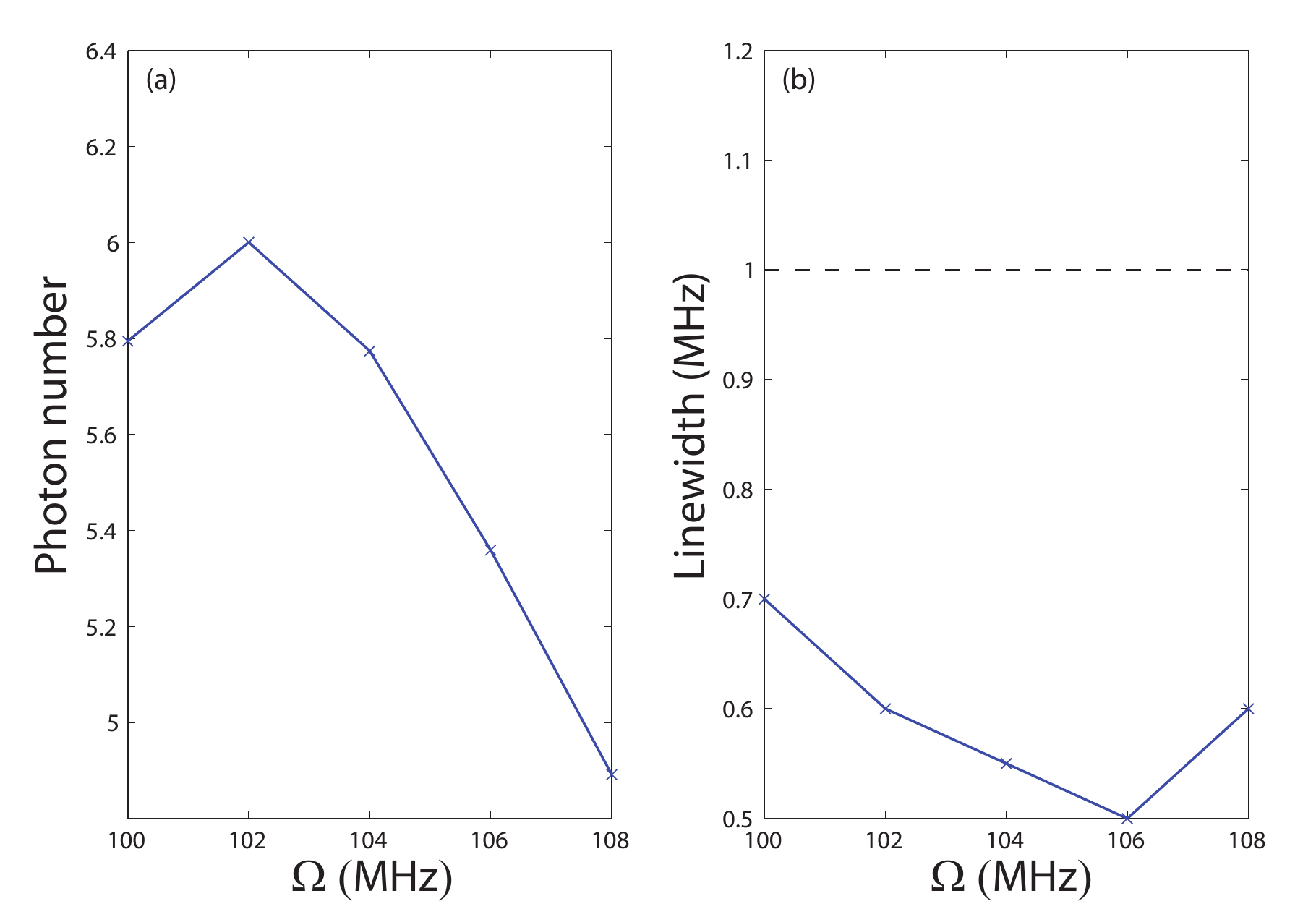}
\caption{Photon number and linewidth of the right mode for varying drive amplitude. Parameters are the same as Fig.~S4 except that $\gamma = 5~\mathrm{MHz}$ instead of $15~\mathrm{MHz}$. The sharpest line does not correspond to the brightest fluorescence. Nevertheless, it is a factor of 2 narrower than the natural linewidth of the bare system. The black dashed line indicates the natural linewidth of the mode ($1~\mathrm{MHz}$).}
{\label{fig:Spectrum Narrowing II}}
\end{figure*} 

\end{document}